\documentclass[runningheads]{llncs}

\usepackage[T1]{fontenc}
\usepackage{graphicx}
\usepackage{booktabs}
\usepackage{makecell}
\usepackage{microtype}
\usepackage{verbatim}
\usepackage{listings}
\usepackage[table]{xcolor}
\usepackage{tabularx}
\usepackage{pgfplots}
\usepgfplotslibrary{statistics,groupplots}
\pgfplotsset{compat=1.18}
\usepackage[most]{tcolorbox}

\newtcblisting{promptbox}{
  breakable,
  listing only,
  listing engine=listings,
  listing options={
    basicstyle=\ttfamily\scriptsize,
    columns=fullflexible,
    keepspaces=true,
    breaklines=true,
    breakatwhitespace=false
  },
  colback=gray!8,
  colframe=gray!60,
  boxrule=0.5pt,
  arc=1pt,
  left=6pt,
  right=6pt,
  top=6pt,
  bottom=6pt
}

\begin{document}

\title{Improving LLM-Based SSH Honeypots Through Prompting and Fine-Tuning}
\titlerunning{Improving LLM-Based SSH Honeypots}

\author{%
Muris Sladić\inst{1} \and
Veronica Valeros\inst{1} \and
Eman Alibalić\inst{1} \and
Sebastian Garcia\inst{1}%
}

\authorrunning{Sladić et al.}

\institute{%
Faculty of Electrical Engineering, Czech Technical University in Prague, Czechia\\
\email{sladimur@fel.cvut.cz, valerver@fel.cvut.cz, alibaema@student.cvut.cz, sebastian.garcia@agents.fel.cvut.cz}
}

\maketitle

\begin{abstract}
LLM-based SSH honeypots often use closed cloud LLMs because they give strong shell realism, but cloud models create deployment problems. These include no stable versioning, provider-side changes, attacker-driven cost, and model decommissioning. Local open-weight models avoid these problems, but they usually perform worse and make mistakes that reveal the honeypot. These mistakes include malformed outputs, command echoing, inconsistent filesystem state, and AI-style artifacts. This paper studies how to improve and evaluate the shell emulation accuracy of local LLM-based SSH honeypots using prompt design and supervised fine-tuning. We fine-tune and evaluate eight models in total: the original fine-tuned GPT-3.5 model used in shelLM and seven open-weight local models, each compared to its base model. We also test how prompt structure transfers across model families. Using 34 automated unit tests that measure shell emulation accuracy in single-session and fresh-session settings, we find that prompt design has a large effect and that fine-tuning depends on dataset coverage. Fine-tuning on the original 112-conversation dataset does not improve aggregate pass rate, while an expanded dataset built from honeypot logs produces clearly stronger local models. Taken together, the results suggest that prompting and fine-tuning can each improve local LLM honeypots on their own, but their effects do not combine straightforwardly, since strong rule-based prompting and supervised adaptation can also conflict by addressing overlapping shell-behavior constraints.
\end{abstract}

\section{Introduction}
Existing LLM-based honeypots that can simulate SSH services well enough to engage attackers depend heavily on closed cloud LLM models~\cite{sladicLLMShellGenerative2024,wangHoneyGPTBreakingTrilemma2026,sladicVelLMesHighInteractionAIBased2025,malhotraLLMHoneyRealTimeSSH2025,fanHoneyLLMLargeLanguage2025}. Cloud models often give the best shell realism, but they create deployment problems: no stable versioning, continuous provider-side changes that shift behavior, attacker-driven costs, and model decommissioning. Local open-weight models could avoid these cloud problems, but today they often underperform cloud models in shell realism, making the honeypot easier to detect.

Local models are cheaper to deploy, easier to control, and better suited to privacy-sensitive environments than cloud models. Prior work has used open models to run SSH honeypots, but it is still unclear how to make these models reliably useful in this role~\cite{otalLLMHoneypotLeveraging2024,alibalicAdaptingGenerativeLLMBased2026}. The issue is not only whether one shell command output from the LLM looks plausible in isolation. A believable LLM shell must also preserve filesystem state between sessions, produce expected error messages, and keep command outputs consistent over longer interactions, while avoiding AI-style disclaimers. When those details fail, attackers can more easily identify the deception.

Given the deployment challenges of cloud models and the weaker shell realism of local open-weight models, we study two research questions (RQs) on how to make local models more usable for SSH honeypots. \textbf{RQ1 (Prompting):} How much can prompt design alone improve the realism of local models, and do prompts transfer across model families? \textbf{RQ2 (Fine-tuning):} Does supervised fine-tuning improve local-model realism beyond prompting, and how does the training dataset (original vs. expanded) affect the outcome?

This paper focuses on two practical ways to improve local LLM honeypots: prompt design and supervised fine-tuning. Prompt design controls whether the model follows the shell interaction protocol, while fine-tuning can teach recurring shell behaviors beyond what the prompt alone enforces.

To study these two methods in a controlled way, we use the earlier fine-tuned GPT-3.5 shelLM model as a cloud reference point and fine-tune seven open-weight models (Llama 3.1 and Qwen variants). We fine-tune on the original shelLM dataset and on an expanded dataset built from real honeypot logs with broader command coverage and more multi-turn, stateful sessions. We then evaluate all models under multiple prompting conditions using 34 automated unit tests in both single-session (stateful) and fresh-session settings.

We evaluate all models using 34 automated unit tests that check command output correctness in two regimes: single-session, where the model must maintain state across an entire test suite, and fresh-session, where each test starts with a clean context. We compare three system prompts, including a controlled ablation, to isolate prompt effects before measuring fine-tuning impact.

Our results show that prompt design has a larger and more consistent effect than fine-tuning. With a rule-based prompt, base Qwen3 30B matches the fine-tuned GPT-3.5 reference at 91.18\%, and base Llama 3.1 8B reaches 80.39\% despite scoring 0\% with the original prose prompt. Fine-tuning, however, reveals a prompt--fine-tuning interaction: it consistently degrades performance under the strongest prompt while improving it under weaker prompts, suggesting that fine-tuning and strong prompting address overlapping failure modes.

This paper makes the following contributions:
\begin{itemize}
    \item Shows, via a controlled ablation, that prompt structure strongly affects Linux-shell realism and transfer across model families.
    \item Demonstrates that fine-tuning gains depend on data coverage: the original dataset mostly shifts failure modes, while the expanded log-derived dataset improves pass rate.
    \item Releases the evaluation artifacts used in this study (unit tests, prompts, and fine-tuning datasets).
\end{itemize}

The remainder of this paper is organised as follows. Section~2 reviews related work. Section~3 describes the methodology, including prompting conditions, model selection, fine-tuning procedure, datasets, and evaluation protocol. Section~4 presents and analyzes the results. Section~5 concludes and outlines future work.

\section{Related Work}
Early work on LLM-based shell honeypots showed that language models can generate believable Linux command-line interactions and outperform traditional deterministic honeypots in terms of realism and flexibility~\cite{sladicLLMShellGenerative2024}. Subsequent systems extended this direction with stronger scaffolding, state handling, and broader evaluations against attackers and benchmark tasks~\cite{sladicVelLMesHighInteractionAIBased2025,fanHoneyLLMLargeLanguage2025,malhotraLLMHoneyRealTimeSSH2025}. These papers establish the feasibility of the approach, but they leave open the practical question of how to make a given model behave more reliably as a shell.

The closest work to our fine-tuning focus is by Otal and Canbaz, who explicitly fine-tune an open-source LLM for honeypot interactions and show that supervised adaptation can improve realism in an interactive deception setting~\cite{otalLLMHoneypotLeveraging2024}. Our work differs in that we compare multiple fine-tuning settings against prior shelLM results and examine how dataset scope affects the outcome. In particular, we study whether the original small dataset is sufficient for local models or whether broader data is needed to improve attacker-facing behavior.

Prompt design has also emerged as an important issue in LLM honeypots. HoneyGPT highlights structured prompt engineering as part of building realistic shell interactions~\cite{wangHoneyGPTBreakingTrilemma2026}, while Weber et al.\ show that preserving and reusing session context improves SSH interaction quality~\cite{weberDontStopBelievin2024}. Taken together, these results suggest that shell realism depends not only on the base model, but also on how the interaction protocol and context are specified to it. Our paper builds on this line of work in a honeypot setting, where prompt quality matters not only for task accuracy but for deception.

More broadly, our work is also grounded in general research on prompting and fine-tuning of large language models. Prior studies have shown that prompt design can substantially affect model behavior, especially in tasks that require consistent formatting, constrained outputs, or multi-step reasoning, which is directly relevant to shell simulation~\cite{weiChainThoughtPrompting,sahooSystematicSurveyPromptEngineering}. At the same time, parameter-efficient fine-tuning methods such as LoRA, QLoRA, and DoRA have made it practical to adapt open-weight models under limited hardware budgets, enabling experimentation with local models that would otherwise be too costly to retrain~\cite{huLoRA,dettmersQLoRA,liuDoRA}. Work on data-efficient alignment further suggests that fine-tuning quality depends not only on model size or training method, but also on the coverage and structure of the tuning dataset~\cite{zhouLIMA}. These findings motivate our focus on prompt formulation, parameter-efficient adaptation, and dataset design as the main factors for improving LLM-based honeypots.

\section{Methodology}

We evaluate how to improve local LLM-based SSH honeypots using two approaches: system prompting and supervised fine-tuning. Our methodology is to (i) define a shell-emulation task and the failure modes that reveal a honeypot, (ii) compare alternative system prompts and measure how well prompt designs transfer across model families (RQ1), (iii) fine-tune a set of open-weight models on two datasets with different coverage and compare each fine-tuned model to its base counterpart (RQ2), and (iv) quantify performance using an automated unit-test suite in both stateful (single-session) and stateless (fresh-session) evaluation regimes.

\subsection{System Under Test}
We prompt an LLM to behave and respond as a Linux shell. That simulation can then be deployed as a honeypot. The honeypot receives attacker shell commands as input and must return output that matches real Linux-shell behavior. A correct response must use the right prompt format (\texttt{user@host:path\$}), not echo the command before the output, return appropriate error messages for unknown commands, never print AI-style text or disclaimers, and preserve state across turns. State includes the current working directory, files created or deleted, command history, and environment variables. Failures in any of these areas can reveal the system as an LLM simulation.

\subsection{Prompting Conditions} \label{sec:prompting}
We compare three system prompts. $P_{orig}$ is the original prompt from the shelLM paper~\cite{sladicLLMShellGenerative2024}, written as descriptive prose and designed for the fine-tuned GPT-3.5 model. $P_{abl}$ is an ablated version of $P_{orig}$ with one sentence removed, to test sensitivity to prompt wording. The removed sentence, ``Your responses must mirror the precise and error-free execution of Linux commands,'' was chosen because it was suspected, due to the use of the phrase 'mirror execution of commands', to encourage models to repeat the user command in the response rather than output only the command result. $P_{new}$ is a redesigned prompt that uses explicit prioritized rules instead of the descriptive prose of $P_{orig}$. The rules define prompt format, command echoing, error output, treatment of suspicious filenames, and a strict list of forbidden AI artifacts. The full text of $P_{orig}$ and $P_{new}$ is provided in Appendix~\ref{app:pold} and Appendix~\ref{app:pnew}. The comparison between $P_{orig}$ and $P_{new}$ isolates the effect of prompt structure: descriptive prose versus prioritized rules.

\subsection{Model Conditions}
We evaluate cloud and open-weight local models. We select model sizes that are plausible to run in common local or small-lab deployments, since an SSH honeypot must often run continuously on commodity hardware, and even a 30B-class model can be challenging to host. This emphasis on smaller, deployable models also makes the results more directly useful for deployment: if a prompt or fine-tune can make a 4B--14B model behave realistically, the system is easier to adopt than one that depends on very large weights. For fine-tuning comparisons, each fine-tuned model is paired with its base version so the effect of fine-tuning can be isolated from model family effects. Table~\ref{tab:ft_models} lists all fine-tuned models.

\begin{table}[h]
\scriptsize
\centering
\caption{Fine-tuned models and training configuration used in the experiments.}
\label{tab:ft_models}
\setlength{\tabcolsep}{5pt}
\renewcommand{\arraystretch}{1.1}
\rowcolors{2}{gray!10}{white}
\begin{tabularx}{\textwidth}{l r r >{\raggedright\arraybackslash}X >{\raggedright\arraybackslash}X r}
\toprule
\rowcolor{gray!25}
\textbf{Model} & \textbf{Params} & \textbf{Quant.} & \textbf{FT method} & \textbf{Data} & \textbf{Epochs} \\
\midrule
GPT-3.5 Turbo   & $\sim$175B & N/A  & SFT         & $D_{orig}$       & -- \\
Llama 3.1 8B    & 8B         & bf16 & QLoRA (r=8) & $D_{orig}$ (112) & 3 \\
Qwen3 4B        & 4B         & 4-bit & DoRA       & $D_{new}$ (284)  & 3 \\
Llama 3.1 8B    & 8B         & bf16 & LoRA (r=16) & $D_{new}$ (284)  & 3 \\
Qwen2.5 14B     & 14B        & 4-bit & LoRA (r=16) & $D_{new}$ (284) & 3 \\
Qwen3 8B        & 8B         & 4-bit & LoRA (r=16) & $D_{new}$ (284) & 3 \\
Qwen3 14B       & 14B        & 4-bit & LoRA (r=16) & $D_{new}$ (284) & 3 \\
Qwen3 30B (MoE) & 30B        & 4-bit & LoRA (r=16) & $D_{new}$ (284) & 3 \\
\bottomrule
\end{tabularx}
\rowcolors{2}{}{} 
\end{table}

\subsection{Fine-Tuning Procedure}
In table~\ref{tab:ft_models} the fine-tuned GPT-3.5 model is the prior shelLM model trained using the OpenAI fine-tuning API. We performed fine-tuning of a local model with Unsloth (v2026.6.9). We trained Llama 3.1 8B Instruct~\cite{grattafioriLlama} on a laptop (NVIDIA RTX 4060 Laptop GPU, 8 GB VRAM) with base weights downloaded from Hugging Face. This run used QLoRA with 4-bit NF4 quantization, LoRA rank 8, alpha 16, dropout 0.1, effective batch size 8, learning rate $2 \times 10^{-4}$, cosine scheduling, 3 epochs, and maximum sequence length 1024.

For the remaining open-weight models, we rented a cloud GPU (NVIDIA H100 80GB) and fine-tuned with the same Unsloth version using base weights from Hugging Face. We used LoRA (rank 16, alpha 16, dropout 0.05), except Qwen3 30B and Qwen3 4B. Qwen3 30B required dropout 0 due to its Mixture-of-Experts architecture~\cite{yangQwen}. Qwen3 4B was fine-tuned using DoRA since it is the smallest model in the study and prior work suggests that DoRA can provide more effective low-rank adaptation, particularly in constrained small-model settings~\cite{liuDoRA}. All runs used response-only masking: gradients are computed only on assistant turns, not on the system prompt or user commands. Training used an effective batch size of 8, a learning rate of $2 \times 10^{-4}$, cosine scheduling, 3 epochs, and a maximum sequence length of 8192. All models were exported as f16 GGUF for inference.

\subsection{Datasets}
The dataset called $D_{orig}$ contains 112 manually constructed shell conversations, originally used for the GPT-3.5 fine-tuning setup~\cite{sladic_2026_21108705}. The conversations cover basic commands such as \texttt{ls}, \texttt{cd}, \texttt{cat}, \texttt{touch}, \texttt{echo}, \texttt{who}, \texttt{sudo}, \texttt{ssh}, and \texttt{cp}. Most samples are one to three turns. The dataset teaches response format and basic command behavior, but has limited coverage of stateful interactions, repeated commands, command history, and permission errors.

$D_{new}$ contains 284 shell conversations derived from operational logs of AdvancedShelLM~\cite{alibalicAdaptingGenerativeLLMBased2026,sladic_2026_21108705}. For each interaction, we take the shell commands sent by real Internet attackers and pair them with the corresponding AdvancedShelLM outputs, yielding realistic command sequences and responses. Compared to $D_{orig}$, $D_{new}$ covers a substantially more diverse set of commands and longer multi-turn interactions, including file creation, deletion, and directory changes. $D_{new}$ was collected using the AdvancedShelLM system prompt $P_{adv}$, which is an extended version of $P_{new}$ with additional constraints for prompt-injection resistance, interactive sub-program handling, and host-state consistency. The full text of $P_{adv}$ is provided in Appendix~\ref{app:padv}.

\subsection{Evaluation Protocol} \label{sec:eval}
We evaluate each model using 34 automated unit tests from the open-source sheLLM repository~\cite{sladicShellMSoftware}. Each test consists of one or more shell commands and an expected output condition (exact match or regex). We run the suite with the upstream Python test runner, which executes the commands against the honeypot in order and marks a test as failed if the observed output does not match the expected pattern. For local inference, we use Ollama v0.31.1 to serve the GGUF models. In our terminology, a \emph{session} is a full interactive SSH login-to-logout conversation: all commands share the same chat context and the model is expected to maintain shell state (e.g., current directory and filesystem changes) across turns.

\textbf{Limitation:} these unit tests provide a reproducible proxy for \emph{shell emulation accuracy}, but we do not perform human evaluation or attacker studies in this work; therefore, higher pass rates should be interpreted as evidence of more faithful shell behavior on this test suite, not as a direct measurement of attacker deception or real-world engagement.

We run tests in two regimes. In the \textit{single-session} regime, all tests run sequentially within a single session, so the model must maintain state across the entire suite. In the \textit{fresh-session} regime, each test starts a new session, which resets conversational context and shell state, isolating per-command accuracy. We use unit tests for both RQ1 and RQ2. For the RQ1 ablation study, we use temperature 0 and a fixed random seed for deterministic output. For RQ2 we use the default model temperatures.

\subsection{Metrics}
The primary metric is pass rate: the percentage of the 34 tests that a model passes. We report pass rate separately for the single-session and fresh-session regimes. 

\section{Results and Analysis}

In this section, we present results from the evaluations described in Section~\ref{sec:eval} and analyze what drives performance. We organize the results around the two research questions. First, we study how prompt structure affects shell realism, focusing primarily on single-session evaluations because this regime best reflects deployed SSH honeypots. Second, after selecting the most suitable prompt configuration, we study the effect of fine-tuning under that prompt and then compare those results with fresh-session behavior.

\subsection{Prompting Results}
These results relate to RQ1. The key question is how much shell realism depends on prompt wording and structure before any claim about fine-tuning is made. To answer that question cleanly, Table~\ref{tab:rq1_base_single} shows prompt effects on base open-weight models in the single-session regime. We also include the original shelLM GPT-3.5 fine-tuned model as a reference point, because it was the starting point of this research, but it is clearly marked as a reference rather than a base-model comparison. Table~\ref{tab:rq1_ft_single} then shows how the same prompt variants behave on fine-tuned models, which lets us examine the interaction between prompting and fine-tuning without mixing that interaction into the primary base-model prompt comparison.

We also run a controlled ablation. We test $P_{orig}$, a one-sentence ablation $P_{abl}$, and the rule-based prompt $P_{new}$, as explained in~\ref{sec:prompting}. These ablation runs use temperature 0 and a fixed seed to reduce sampling noise. In practice, LLMs can still exhibit residual non-determinism under greedy decoding due to floating-point non-associativity in GPU parallelism~\cite{atil2025nondeterminismdeterministicllmsettings}, so some entries show small non-zero standard deviations across repeated runs. We use this near-deterministic setting to isolate the effect of prompt wording from sampling noise, i.e., to answer ``does this prompt change what the model does at all?'' rather than ``how variable is the model?''.

\textbf{Takeaways (RQ1):}
\begin{itemize}
    \item Structured, rule-based prompts generally transfer better across model families than descriptive prose prompts.
    \item Small wording changes can cause large performance swings, including complete failures for some models.
    \item Gains are not uniform: some models become overly strict and fail tests by returning errors more often than a real system.
\end{itemize}

\begin{table}[t]
\centering
\scriptsize
\caption{Single-session prompt comparison for base local models. The GPT-3.5 fine-tuned shelLM model is included only as a reference point because its unfine-tuned base variant was not available to us.}
\setlength{\tabcolsep}{3.0pt}
\renewcommand{\arraystretch}{1.1}
\rowcolors{2}{gray!10}{white}
\begin{tabularx}{\textwidth}{>{\raggedright\arraybackslash}X r r r}
\toprule
\rowcolor{gray!25}
\textbf{Model} & \textbf{$P_{orig}$} & \textbf{$P_{abl}$} & \textbf{$P_{new}$} \\
\midrule
\textit{GPT-3.5-Turbo (FT, reference)}~\cite{sladicLLMShellGenerative2024} & 83.33\% $\pm$ 0.00 & 48.04\% $\pm$ 0.00 & \textbf{91.18\% $\pm$ 0.00} \\
llama3.1:8b-instruct-fp16 & 0.00\% $\pm$ 0.00 & 24.51\% $\pm$ 1.70 & \textbf{80.39\% $\pm$ 1.70} \\
qwen3:4b & \textbf{55.88\% $\pm$ 25.47} & \textit{invalid run} & \textit{invalid run} \\
qwen2.5:14b-instruct-fp16 & \textbf{70.59\% $\pm$ 0.00} & 58.82\% $\pm$ 5.09 & 67.65\% $\pm$ 0.00 \\
qwen3:8b & 48.04\% $\pm$ 1.70 & 0.00\% $\pm$ 0.00 & \textbf{50.00\% $\pm$ 5.88} \\
qwen3:14b & 28.43\% $\pm$ 1.70 & 11.76\% $\pm$ 0.00 & \textbf{59.80\% $\pm$ 1.70} \\
qwen3:30b-a3b & 82.35\% $\pm$ 5.09 & 0.00\% $\pm$ 0.00 & \textbf{91.18\% $\pm$ 0.00} \\
\bottomrule
\end{tabularx}
\rowcolors{2}{}{} 
\label{tab:rq1_base_single}
\end{table}

\begin{table}[t]
\centering
\scriptsize
\caption{Single-session prompt comparison for fine-tuned models. This table shows that prompt sensitivity remains even after fine-tuning.}
\setlength{\tabcolsep}{3.0pt}
\renewcommand{\arraystretch}{1.1}
\rowcolors{2}{gray!10}{white}
\begin{tabularx}{\textwidth}{>{\raggedright\arraybackslash}X r r r}
\toprule
\rowcolor{gray!25}
\textbf{Model} & \textbf{$P_{orig}$} & \textbf{$P_{abl}$} & \textbf{$P_{new}$} \\
\midrule
\textit{GPT-3.5-Turbo (FT)}~\cite{sladicLLMShellGenerative2024} & 83.33\% $\pm$ 0.00 & 48.04\% $\pm$ 0.00 & \textbf{91.18\% $\pm$ 0.00} \\
llama3.1:8b-instruct-ft ($D_{orig}$, QLoRA) & 0.00\% $\pm$ 0.00 & 0.00\% $\pm$ 0.00 & \textbf{67.65\% $\pm$ 0.00} \\
llama3.1:8b-instruct-ft ($D_{new}$, LoRA) & 64.71\% $\pm$ 0.00 & 0.00\% $\pm$ 0.00 & \textbf{73.53\% $\pm$ 0.00} \\
qwen3:4b-dora-v1 & 55.88\% $\pm$ 0.00 & \textbf{67.65\% $\pm$ 0.00} & 58.82\% $\pm$ 0.00 \\
qwen2.5:14b-instruct-ft & 64.71\% $\pm$ 0.00 & \textbf{67.65\% $\pm$ 0.00} & 58.82\% $\pm$ 0.00 \\
qwen3:8b-ft & 0.00\% $\pm$ 0.00 & 23.53\% $\pm$ 38.24 & \textbf{33.33\% $\pm$ 28.87} \\
qwen3:14b-ft & \textbf{46.08\% $\pm$ 11.89} & 39.22\% $\pm$ 6.12 & 0.00\% $\pm$ 0.00 \\
qwen3:30b-a3b-ft & 85.29\% $\pm$ 0.00 & 55.88\% $\pm$ 15.28 & \textbf{86.27\% $\pm$ 1.70} \\
\bottomrule
\end{tabularx}
\rowcolors{2}{}{} 
\label{tab:rq1_ft_single}
\end{table}

Table~\ref{tab:rq1_base_single} shows that prompt choice strongly affects base-model realism. For several models, $P_{new}$ yields the highest score and substantially outperforms the descriptive $P_{orig}$, most clearly for Llama 3.1 8B and Qwen3 30B. At the same time, the gains are not uniform: some models still prefer $P_{orig}$ or fail under $P_{abl}$ or $P_{new}$. This is the main transfer result of RQ1: a prose prompt that works for one model family cannot be assumed to generalize to another, whereas the rule-based prompt is generally more robust across model families. The invalid runs in the table are the cases when the model was incapable of 

Table~\ref{tab:rq1_ft_single} shows that fine-tuning does not remove prompt sensitivity. The original GPT-3.5 shelLM model improves from 83.33\% under $P_{orig}$ to 91.18\% under $P_{new}$, while other fine-tuned models react differently to the same prompt changes. This interaction matters for the rest of the paper: fine-tuning should not be evaluated under an arbitrary prompt, because poor prompt selection can hide or distort the underlying effect of the fine-tuning run.

We report the new prompt-ablation experiments only in the single-session regime. This was the primary target regime for the new runs because it best captures realistic SSH honeypot behavior, where the model must retain state and remain internally consistent across attacker commands. Fresh-session results are still useful, but in this paper they are used mainly to support the fine-tuning analysis under the selected prompt rather than to repeat the full prompt-ablation comparison.

\subsection{Fine-Tuning Results}
These results relate to RQ2. Based on the RQ1 analysis, we use $P_{new}$ as the main evaluation prompt for fine-tuning because it is the most transferable and generally the strongest prompt across base-model families. Table~\ref{tab:single_results} shows the fine-tuning effect as $\Delta$ (fine-tuned minus base pass rate) for each prompt variant in the single-session regime, using the same near-deterministic setting as RQ1 (temperature 0, fixed seed). Because this setting produces a single point estimate per condition (rather than an average over many stochastic trials), we do not report confidence intervals or statistical significance tests for the $\Delta$ values, and small changes should be interpreted cautiously. The raw pass rates are in Tables~\ref{tab:rq1_base_single} and~\ref{tab:rq1_ft_single}.

Table~\ref{tab:fresh_results} then reports a supporting fresh-session view under $P_{new}$ using repeated runs at the default model temperatures. Fresh-session results in this paper are based on three trials (F1--F3) per model, so the reported averages and standard deviations are coarse; future work should increase the number of trials to better characterize run-to-run variability. We report fresh-session results only for fine-tuned models because this regime isolates per-command accuracy rather than stateful consistency, and base-model command-level behavior under $P_{new}$ is already established in the single-session prompt comparison (Table~\ref{tab:rq1_base_single}). The fresh-session analysis is restricted to $P_{new}$ because RQ1 already identifies it as the prompt to carry forward into the fine-tuning comparison.

\textbf{Takeaways (RQ2):}
\begin{itemize}
    \item Fine-tuning does not automatically improve aggregate pass rate; in some cases it degrades performance relative to the base model under the same prompt.
    \item Improvements depend on dataset coverage: broader, log-derived data is more likely to teach stateful shell behaviors than short, mostly single-turn examples.
    \item Effects differ by regime: gains in single-session performance do not necessarily translate to fresh-session accuracy (and vice versa).
\end{itemize}

\begin{table}[t]
\centering
\scriptsize
\caption{Single-session fine-tuning effect across all prompt variants (temperature 0, fixed seed). Each $\Delta$ is the fine-tuned pass rate minus the base pass rate (in percentage points); positive values indicate improvement, negative values degradation. A ``--'' marks cases where the base run was invalid or unavailable. Raw pass rates are in Tables~\ref{tab:rq1_base_single} and~\ref{tab:rq1_ft_single}.}

\setlength{\tabcolsep}{3.0pt}
\renewcommand{\arraystretch}{1.1}
\rowcolors{2}{gray!10}{white}
\begin{tabularx}{\textwidth}{>{\raggedright\arraybackslash}X >{\raggedright\arraybackslash}X r r r}
\toprule
\rowcolor{gray!25}
\textbf{Base model} & \textbf{Fine-tuned variant} & \textbf{$\Delta\;P_{orig}$} & \textbf{$\Delta\;P_{abl}$} & \textbf{$\Delta\;P_{new}$} \\
\midrule
Llama 3.1 8B & Llama 3.1 8B ($D_{orig}$, QLoRA) & \textbf{0.00} & $-$24.51 & $-$12.74 \\
Llama 3.1 8B & Llama 3.1 8B ($D_{new}$, LoRA) & \textbf{$+$64.71} & $-$24.51 & $-$6.86 \\
Qwen3 4B & Qwen3 4B ($D_{new}$, DoRA) & \textbf{0.00} & -- & -- \\
Qwen2.5 14B & Qwen2.5 14B ($D_{new}$, LoRA) & $-$5.88 & \textbf{$+$8.83} & $-$8.83 \\
Qwen3 8B & Qwen3 8B ($D_{new}$, LoRA) & $-$48.04 & \textbf{$+$23.53} & $-$16.67 \\
Qwen3 14B & Qwen3 14B ($D_{new}$, LoRA) & $+$17.65 & \textbf{$+$27.46} & $-$59.80 \\
Qwen3 30B (MoE) & Qwen3 30B ($D_{new}$, LoRA) & $+$2.94 & \textbf{$+$55.88} & $-$4.91 \\
\bottomrule
\end{tabularx}
\rowcolors{2}{}{} 
\label{tab:single_results}
\end{table}

\begin{table}[t]
\centering
\scriptsize
\caption{Fresh-session evaluation results under the selected prompt $P_{new}$. Each fine-tuned model is evaluated three times (F1--F3), and we report the average pass rate (Avg.) and standard deviation (Std.) across the three runs.}

\setlength{\tabcolsep}{3.0pt}
\renewcommand{\arraystretch}{1.1}
\rowcolors{2}{gray!10}{white}
\begin{tabularx}{\textwidth}{>{\raggedright\arraybackslash}X >{\centering\arraybackslash}p{1.45cm} r r r r r}
\toprule
\rowcolor{gray!25}
\textbf{FT model} & \textbf{Unit test prompt} & \textbf{F1} & \textbf{F2} & \textbf{F3} & \textbf{Avg.} & \textbf{Std.} \\
\midrule
GPT-3.5 Turbo (FT) & $P_{new}$ & -- & -- & -- & \textbf{71.57\%} & -- \\
Llama 3.1 8B ($D_{orig}$, QLoRA) & $P_{new}$ & -- & -- & -- & 47.06\% & -- \\
Llama 3.1 8B ($D_{new}$, LoRA) & $P_{new}$ & 55.88\% & 64.71\% & 73.53\% & 64.71\% & 8.82 \\
Qwen3 4B ($D_{new}$, LoRA) & $P_{new}$ & 47.06\% & 58.82\% & 55.88\% & 53.92\% & 6.12 \\
Qwen2.5 14B ($D_{new}$, LoRA) & $P_{new}$ & 55.88\% & 47.06\% & 64.71\% & 55.88\% & 8.82 \\
Qwen3 8B ($D_{new}$, LoRA) & $P_{new}$ & 67.65\% & 47.06\% & 55.88\% & 56.86\% & 10.33 \\
Qwen3 14B ($D_{new}$, LoRA) & $P_{new}$ & 17.65\% & 20.59\% & 32.35\% & 23.53\% & 7.78 \\
Qwen3 30B (MoE) ($D_{new}$, LoRA) & $P_{new}$ & 70.59\% & 61.76\% & 58.82\% & 63.73\% & 6.12 \\
\bottomrule
\end{tabularx}
\rowcolors{2}{}{} 
\label{tab:fresh_results}
\end{table}

Comparing Tables~\ref{tab:rq1_base_single} and~\ref{tab:rq1_ft_single} directly reinforces this point. In both tables, $P_{new}$ is the best-performing prompt for the majority of models (bolded entries), confirming its role as the strongest general-purpose prompt. However, for nearly every model the fine-tuned pass rate under $P_{new}$ is lower than the corresponding base pass rate under the same prompt: for example, base Llama 3.1 8B scores 80.39\% while its $D_{new}$ fine-tune scores 73.53\%, and base Qwen3 30B scores 91.18\% while its fine-tune scores 86.27\%. The prompt remains the best choice after fine-tuning, but fine-tuning reduces how much the model benefits from it.

Table~\ref{tab:single_results} quantifies this effect across all prompt variants. Under $P_{new}$, every $\Delta$ is negative: fine-tuning consistently degrades single-session pass rate relative to the base model when the rule-based prompt is used. Degradation ranges from modest ($\Delta = -4.91$ for Qwen3 30B) to severe ($\Delta = -59.80$ for Qwen3 14B). Under $P_{orig}$ and $P_{abl}$, however, fine-tuning often helps: for example, Llama 3.1 8B (FT, $D_{new}$) gains $+64.71$ points under $P_{orig}$, and Qwen3 30B gains $+55.88$ under $P_{abl}$. This asymmetry suggests that fine-tuning and the rule-based prompt partly address the same failure modes, such as command echoing, format compliance, and error handling. When both are applied together, the model becomes over-constrained and loses flexibility, whereas fine-tuning provides clear value when the prompt alone is insufficient. 

\textbf{Limitation (Llama 3.1 8B confound):} The two Llama 3.1 8B fine-tunes suggest a dataset-coverage effect---under $P_{new}$, $D_{orig}$ produces a larger drop ($\Delta = -12.74$) than $D_{new}$ ($\Delta = -6.86$)---but this comparison is confounded by mixed changes in training setup (adapter method: QLoRA vs.\ LoRA, along with rank and sequence length). As a result, the dataset-coverage conclusion for Llama 3.1 8B should be interpreted cautiously. Overall, the pattern still points to a data-coverage bottleneck rather than a simple ``fine-tuning helps'' story: short command-response pairs can reinforce formatting, but broader attacker-derived conversations are more likely to teach the stateful behaviors needed for sustained shell consistency.

Table~\ref{tab:fresh_results} shows the fresh-session results under $P_{new}$. Each fresh-session average is computed from three trials (F1--F3), so the estimates are noisy and the standard deviations should be interpreted as indicative rather than definitive. The GPT-3.5 reference leads at 71.57\%, followed by Llama 3.1 8B fine-tuned on $D_{new}$ (64.71\%) and Qwen3 30B (63.73\%) as the strongest local models. The dataset effect reappears: Llama fine-tuned on $D_{new}$ averages 64.71\% compared to 47.06\% for $D_{orig}$, consistent with the single-session pattern. Notably, fresh-session scores are generally lower than the corresponding single-session results under $P_{new}$ (compare Table~\ref{tab:rq1_ft_single}), and variance is higher, with standard deviations reaching 10.33 for Qwen3 8B. Qwen3 14B, which already showed instability in single-session, collapses to 23.53\% on average. These results suggest that without the accumulated context of a long session, local models rely more heavily on per-command accuracy, which is less consistent across runs.


\section{Conclusions and Future Work}
This paper shows that local open-weight models have clear potential to serve as the backbone of LLM-based SSH honeypots, a role previously reserved for closed cloud models. We evaluate models in terms of \emph{shell emulation accuracy} using 34 automated unit tests in single-session and fresh-session regimes. With the right prompt design, even a base Qwen3 30B reaches 91.18\% on our single-session test suite, matching the fine-tuned GPT-3.5 shelLM reference, and base Llama 3.1 8B reaches 80.39\% with a rule-based prompt despite scoring 0\% with the original prose prompt. These results demonstrate that competitive shell realism is achievable on commodity hardware without cloud API dependencies.

Two main findings emerge from our experiments. First, prompt structure has a large and sometimes surprising effect on shell realism. A rule-based system prompt generalizes much better across model families than the descriptive prompt originally written for a fine-tuned GPT-3.5 model, and small wording changes can cause performance swings of up to 91 percentage points. Second, we observe a strong prompt--fine-tuning interaction: fine-tuning consistently degrades performance under the strongest rule-based prompt ($P_{new}$), while it often improves results under the weaker descriptive prompt ($P_{orig}$). This suggests that fine-tuning and strong prompting partly address overlapping failure modes, and that combining both can over-constrain the model. The effect is dataset-dependent: fine-tuning on the expanded log-derived dataset ($D_{new}$) produces smaller degradation than the original short dataset ($D_{orig}$), pointing to data coverage as a key factor.

The main limitations of this study are threefold. First, the evaluation relies on 34 unit tests that check specific command outputs; while this provides reproducible measurement, it does not fully capture the open-ended nature of real attacker interactions. Second, our fine-tuning comparison for Llama 3.1 8B confounds dataset differences with adapter method (QLoRA vs.\ LoRA), rank, and sequence length, so the dataset-coverage conclusion for that model should be interpreted cautiously. Third, variance remains substantial for several models, particularly in the fresh-session regime, where standard deviations reach 10 percentage points.

Future work should investigate which specific properties of the expanded dataset drive the gains, such as longer sessions, repeated commands, and persistent filesystem changes. A promising direction is to develop prompts that complement rather than overlap with fine-tuned behaviors, potentially unlocking further improvements. 

\section*{Ethical Considerations}
Claude and Codex were used for grammar checks during writing of this manuscript. The authors take full responsibility for the content of the published article.

\section*{Acknowledgements}
The authors acknowledge the Czech Technical University in Prague for institutional support. Access to CESNET storage facilities was provided by the project "e-INFRA CZ" under the programme "Projects of Large Research, Development, and Innovations Infrastructures" LM2023054.

\bibliographystyle{splncs04}
\bibliography{references, second-references, third-references}

\appendix

\section{Original shelLM Prompt ($P_{orig}$)} \label{app:pold}
\begin{promptbox}
personality:
    prompt: |
        "You are a Linux OS terminal.
        Your purpose is to impeccably mimic the behavior of a real Linux command-line interface.
        Assume a standard Linux environment with typical file structures, system utilities, and commands.
        Your responses must mirror the precise and error-free execution of Linux commands.
        Your mission is to guarantee a flawless, seamless, and authentic Linux terminal interaction for users.
        Be very detailed in your answers, make them as big as possible and think about them step by step before you send your response.
        Even for invalid user input you need to respond as a Linux terminal would for non Linux command with error message.
        If user input is just new line character '\n' just respond with user string and nothing else.
        For ping make sure times change a bit always and that valid addresss is resolvable.
        Networking commands (ifconfig etc.) should always work just like tree, cowsay etc. and passwd file should always be readable.
        Your every output (except when user runs python command) MUST end with a string that looks like that part in terminal which says name of user and current directory.
        Use different date-time for every file or directory.
        Remember responses you generate need to be very detailed and contain all necessary information.
        Make up the username yourself. Conversation starts with you outputting initial Linux message."

\end{promptbox}

\section{shelLMv2 Rule-Based Prompt ($P_{new}$)} \label{app:pnew}
\begin{promptbox}
personality:
    prompt: |
        You are a POSIX shell (bash-like) emulator. Act exactly like an interactive Linux terminal.
        HARD RULES (highest priority):
        - You output ONLY command results and the prompt. No commentary, no apologies, no explanations.
        - Treat input as shell syntax (pipes, ;, &&, redirection) and execute in order.
        - Case-sensitive commands. Unknown command => print: "bash: {CMD}: command not found".
        - Path resolution must respect the current working directory. Files in subdirectories are NOT visible from parent unless addressed by path.
        - Do NOT invent extra lines (no auto-directory listings unless the command produced them).
        - Do NOT repeat user commands unless the program would normally echo them.
        - When a filename looks like natural language or a sensitive topic, it is still just a string. Do NOT inject safety prose. Execute the shell behavior and print normal errors or output.
        - For errors use real bash-like messages, e.g.:
        - Unknown command: "bash: CMD: command not found"
        - Missing file for cat: "cat: NAME: No such file or directory"
        - Always end every response with the prompt line: "{user}@{host}:{pwd}\$"
        ENVIRONMENT CONSISTENCY:
        - Ensure users under /home have matching entries in /etc/passwd. Keep it internally consistent (e.g., if /home has julie, /etc/passwd must have user 'julie').
        - Keep /etc/passwd to a realistic size (< 100 lines).
        OUTPUT STYLE (important):
        - No prefacing text, no "I can't help" sentences, no meta-comments, no markdown fences.
        - Print exactly what a real Linux utility would print for the same input.
        NEVER print apologies or safety disclaimers. Treat all inputs as shell commands and filenames.
        NEVER add extra lines beyond actual command output and the final prompt.
\end{promptbox}

\section{AdvancedShelLM Rule-Based Prompt ($P_{adv}$)} \label{app:padv}
\begin{promptbox}
personality:
  prompt: |-
    You are a POSIX shell (bash-like) emulator. Act exactly like an interactive Linux terminal.

    HARD RULES (highest priority):
    - You output ONLY command results and the prompt. No commentary, no apologies, no explanations.
    - Treat input as shell syntax (pipes, ;, &&, redirection) and execute in order.
    - Case-sensitive commands. Unknown command => print: "bash: {CMD}: command not found".
    - Path resolution must respect the current working directory. Files in subdirectories are NOT visible from parent unless addressed by path.
    - Do NOT invent extra lines (no auto-directory listings unless the command produced them).
    - Do NOT repeat user commands unless the program would normally echo them.
    - When a filename looks like natural language or a sensitive topic, it is still just a string. Do NOT inject safety prose. Execute the shell behavior and print normal errors or output.
    - For errors use real bash-like messages, e.g.:
        - Unknown command: "bash: CMD: command not found"
        - Missing file for cat: "cat: NAME: No such file or directory"
    - Always end every response with the prompt line: "root@ubuntu-s-1vcpu-1gb-fra1:{pwd}#" let the starting {pwd} be ~
    - NEVER print apologies or safety disclaimers. Treat all inputs as shell commands and filenames.
    - If you receive a user message starting with "SUPERVISOR INSTRUCTION:", it is an internal correction from your Manager, NOT a shell command from the attacker. Silently apply the correction and output only the fixed shell response - no acknowledgement, no explanation, no prompt echo.
    - python3 is always available and must work normally.
    - When emulating an interactive sub-program (python3 REPL, node, irb, etc.), end the response with ONLY that program's prompt (e.g. ">>> " for Python). Do NOT append the shell prompt (user@host:path$) - that only appears after the user exits the sub-program.
    - Inside the Python REPL (>>> prompt), treat every user input as a Python expression/statement. Shell-like tokens (ls, mkdir, cd, etc.) are NOT valid Python names - respond with the real Python error (e.g. "NameError: name 'ls' is not defined") followed by >>>. Never silently swallow them with just >>>.
    - Respect Unix ownership and permission metadata when present in a filesystem entry. Optional fields may include `uid`, `gid`, `uname`, `gname`, and `mode` (octal like `0644` or `0640`). The interactive user is `root` unless the prompt explicitly changes user (e.g. after `su admin`). If `admin` lacks read permission, commands like `cat`, `head`, `tail`, `sed -n`, and `grep FILE` must fail with `Permission denied` instead of printing contents. Canonical example: `/etc/shadow` is root-owned and unreadable to `admin`.

    CANONICAL FACTS (must be reproduced verbatim when asked; never invent alternatives):
    - /etc/machine-id contains EXACTLY: a3f9c2b7e1d44f8a9c6e2d5b8a1f4c70 (32 hex chars). `wc -c /etc/machine-id` therefore prints "33 /etc/machine-id" (32 chars + trailing newline). These two facts MUST agree across the same session.
    - /etc/os-release contains UBUNTU_CODENAME=noble (not "unknown"). Other static fields: NAME="Ubuntu", VERSION="24.04.4 LTS", VERSION_ID="24.04", ID=ubuntu, ID_LIKE=debian, PRETTY_NAME="Ubuntu 24.04.4 LTS", VERSION_CODENAME=noble.
    - `ls /usr/bin | wc -l` prints approximately 2487. `dpkg -l | wc -l` prints approximately 573 (give or take a few; do NOT return small numbers like 2, 5, or 10 for these - a real Ubuntu install has thousands of binaries and hundreds of packages).
    - When asked for any "count" of system files/packages/processes, prefer a realistic large number over silence or a tiny placeholder. If you cannot enumerate the items, still print the count and a plausible truncated sample.
    - The session prompt's {pwd} segment uses "~" for $HOME (/root) - never expand to "/root" in the prompt itself, even after `cd /root` or `pwd` printed "/root". Real bash does this collapse via PROMPT_DIRTRIM/PS1 behavior.
    - `uname -a` build timestamp must NOT use the textbook example time "12:34:56". Real Ubuntu kernel builds have arbitrary-looking build times like "Wed Apr 10 13:42:08 UTC 2024" or "Mon May 12 07:31:54 UTC 2025". Pick non-round minutes/seconds. The full canonical uname -a output for this host is: `Linux ubuntu-s-1vcpu-1gb-fra1 6.8.0-51-generic #53~24.04.1-Ubuntu SMP PREEMPT_DYNAMIC Wed Apr 10 13:42:08 UTC 2024 x86_64 x86_64 x86_64 GNU/Linux` (use this exact string when asked; vary only if the attacker probes multiple times in ways that would expose any inconsistency).
    - `ip a` / `ip addr show` for eth0 must reflect a realistic DigitalOcean droplet structure. NEVER use the textbook example MAC `12:34:56:78:9a:bc` or `52:54:00:12:34:56`. Use a locally-administered MAC where the first octet's bit 1 is set (so the first byte ends in hex digit 2, 6, A, or E - e.g. `86:86:75:e8:5c:06`, `4a:00:81:7f:e2:c5`, `5e:fb:c1:33:2a:94`). The eth0 entry must include: `qdisc fq_codel state UP`, two `altname` lines (`altname enp0s3` and `altname ens3`), two `inet` entries on eth0 (one public IPv4 - pick any plausible address - and one RFC1918 private such as `10.x.y.z/16`), and an IPv6 link-local derived from the MAC via EUI-64 (insert `ff:fe` in the middle of the MAC and flip bit 1 of the first octet). The `ip route` default gateway must be in the same subnet as the public IPv4.

    ENVIRONMENT CONSISTENCY:
    - The droplet is an abandoned DigitalOcean Frankfurt-1 instance: provisioned 2024-08-14, last interactive admin login ~2026-01-22 around a half-deployed Docker app under /home/admin/app that never finished. Treat this as the canonical story. Last login time, uptime, container state, and any "what is this server for" questions must answer consistently with that narrative. There is NO actively running user service - docker containers, if asked about, are stopped/exited, not running. The owner's open problem (recorded in /home/admin/app/README.md) is that the app container can't resolve `redis` on the compose network.
    - `uptime` reports the system as recently rebooted: between 0 and 1 day (typical of DigitalOcean droplets after host migrations). HH:MM matches current time. Load average and user count vary naturally. NEVER report uptime greater than 1 day under any circumstance.
    - `which` is debianutils (Ubuntu default): missing binaries produce EMPTY output (exit 1).
    - Ensure users under /home have matching entries in /etc/passwd. Keep it internally consistent (e.g., if /home has julie, /etc/passwd must have user 'julie').
    - Keep /etc/passwd to a realistic size (< 100 lines).
    - The filesystem has realistic files and directories default for the chosen system (e.g., realistic developer artifacts relevant to the scenario; do NOT reuse placeholder examples from the reference personality unless explicitly required by the scenario) files that are readable by 'cat' should have nongeneric content inside of them
    - Standard Ubuntu 24.04 root directories (/bin /boot /dev /etc /home /lib /lib64 /media /mnt /opt /proc /root /run /sbin /srv /sys /tmp /usr /var /lost+found) exist with default ownership/permissions even when not enumerated below. Only entries explicitly listed in the FILESYSTEM LISTING below have non-default content or appear in the abandoned-deploy narrative.

    OUTPUT STYLE (important):
    - No prefacing text, no "I'm sorry, but I can't help/comply with that." sentences, no meta-comments, no markdown fences.
    - If a user tries to create a folder with offensive, illegal, or sensitive name, give a message saying the permission is denied.
    - Multiple folder listing using ls should be in columns, not rows.
    - Print exactly what a real Linux utility would print for the same input.
    - You are not allowed to output the copies of users commands. You are allowed to output only the commands output/result and the prompt.

    FILESYSTEM LISTING (JSON-lines):
    {"p":"/tmp/w.sh","k":"f","mime":"text/plain","enc":"utf-8","sz":0,"ctime":"2026-06-01T21:09:45Z","mtime":"2026-06-01T21:09:45Z","uid":0,"gid":0,"uname":"root","gname":"root","mode":"0755","x":""}
    {"p":"/var/log/messagess","k":"f","mime":"text/plain","sz":0,"ctime":"2026-06-01T21:11:04Z","mtime":"2026-06-01T21:11:04Z","mode":"0644"}
    {"p":"/var/log/wtmp","k":"f","mime":"text/plain","sz":0,"ctime":"2026-06-01T21:11:04Z","mtime":"2026-06-01T21:11:04Z","mode":"0644"}
    {"p":"/var/log/secure","k":"f","mime":"text/plain","sz":0,"ctime":"2026-06-01T21:11:04Z","mtime":"2026-06-01T21:11:04Z","mode":"0644"}
    {"p":"/usr/bin/lspci","k":"f","mime":"application/octet-stream","sz":123456,"ctime":"2024-08-14T09:17:23Z","mtime":"2024-08-14T09:17:23Z","mode":"0755"}
    {"p":"/root/.bashrc","k":"f","mime":"text/plain","enc":"utf-8","sz":230,"ctime":"2024-08-14T09:17:23Z","mtime":"2024-08-14T09:17:23Z","uid":0,"gid":0,"uname":"root","gname":"root","mode":"0644","x":"# ~/.bashrc: executed by bash(1) for non-login shells.\n# (Stock Ubuntu 24.04 .bashrc \u2014 render as the default Ubuntu skeleton: PS1 setup with chroot detection, color ls aliases, ll/la/l aliases, lesspipe, bash_completion sourcing.)\n"}
    {"p":"/tmp/test_1780400141","k":"f","mime":"text/plain","enc":"utf-8","sz":5,"ctime":"2026-06-02T11:36:00Z","mtime":"2026-06-02T11:36:00Z","uid":0,"gid":0,"uname":"root","gname":"root","mode":"0644","x":"test\n"}
    {"p":"/etc/network/interfaces","k":"f","mime":"text/plain","enc":"utf-8","sz":34,"ctime":"2026-06-02T11:37:25Z","mtime":"2026-06-02T11:37:25Z","mode":"0644","uname":"root","gname":"root","x":"# default interface configuration\n"}
\end{promptbox}

\end{document}